\documentstyle[12pt]{article}
\def\half{{1\over 2}}

\def\gsim{\mathrel{\rlap{\raise.5ex\hbox{$>$}}{\lower.5ex\hbox{$\sim$}}}}
\def\lsim{\mathrel{\rlap{\raise.5ex\hbox{$<$}}{\lower.5ex\hbox{$\sim$}}}}
\def\plt{{\it Phys. Lett. }}
\def\np{{\it Nucl. Phys. }}
\def\pr{{\it Phys. Rev. }}
\def\prl{{\it Phys. Rev. Lett. }}
\begin{document}
\begin{titlepage}
\typeout{ *** Title page *** }
\vspace*{-.7in}
\hspace*{5.2in}
\vbox{\hbox{UIOWA-93-11}}
\vspace{.5in}
\begin{center}
\renewcommand{\thefootnote}{\fnsymbol{footnote}}
{\LARGE\bf Enhancing the Higgs signal in $pp\rightarrow ZZ\rightarrow
\ell^+\ell^-\nu\bar{\nu}$\footnote{To be published in the proceedings
of the {\em Workshop on Physics at Current Accelerators and the
Supercollider}, Argonne National Laboratory, 2-5 June, 1993 {\em eds.}
J.L.Hewett, L.E.Price and A.R.White.}
\\}
\bigskip\medskip
\large{\bf M.J.Duncan\footnote{Email: duncan@hepmips.physics.uiowa.edu}}\\
and\\
\large{\bf M.H.Reno\footnote{Email: reno@hephp1.physics.uiowa.edu}}\\
\bigskip
{\sl Department of Physics and Astronomy,\\ University of Iowa,\\
Iowa City, IA 52242}\\
\bigskip
\end{center}
\vfill
\begin{abstract}
One of the possible modes suggested for detecting the Higgs particle is
$pp\rightarrow ZZ\rightarrow\ell^+\ell^-\nu\bar{\nu}$, where the Higgs appears
as a resonance on a Jacobian background. Unfortunately there are QCD background
processes which mimic the final state and we are obliged to impose stringent
kinematic cuts to remove them. However, in doing so we also remove a
significant
fraction of the signal. In this report we suggest a method, based on examining
the distribution of the final state charged leptons, which could allow us to
relax our cuts and thus salvage the signal.
\end{abstract}
\vfill
\end{titlepage}
\setcounter{footnote}{0}
\renewcommand{\thefootnote}{\fnsymbol{footnote}}
\baselineskip = 18pt plus 0.2pt minus 0.1pt    
\typeout{*** Text ***}
There are many methods suggested for detecting a Higgs particle of mass
above 400 GeV at the SSC. The cleanest method by far is searching for a
resonance in the invariant mass distribution of the four final state leptons
in the `gold-plated' decay $pp\rightarrow ZZ\rightarrow\ell_1^+\ell_1^-
\ell_2^+\ell_2^-$. However, due to the small branching fraction for $Z$
decaying
to charged leptons, the number of events is not expected to be large given the
design luminosity of the accelerator, and so there is a possibility that a
heavy Higgs could escape detection because of low statistics. With this in
mind it was suggested\cite{Cahn} that the mode
$pp\rightarrow ZZ\rightarrow\ell^+\ell^-\nu\bar{\nu}$ be considered since
the cross section is a factor of six higher.

The signature for these events is a pair of charged leptons with invariant mass
$m_Z$, and a transverse momentum imbalance ascribed to the neutrinos.
The kinematic variable used to
characterize the cross section is the transverse mass,
\begin{equation}
m_T^2=\left[ ({\bf p}_T^2+m_Z^2)^\half + (\not\!{\bf
p}_T^2+m_Z^2)^\half\right]^2
-({\bf p}_T+\not\!{\bf p}_T)^2
\end{equation}
In Fig. 1 we plot the differential cross sections for $pp\rightarrow ZZ$
as a function of the invariant mass of the $Z$ pair, and as a function of the
transverse mass. What is clear from these figures is that the Higgs is
more pronounced in the transverse mass distribution. This is because in the
LAB frame the heavy Higgs is produced with very little momentum
and then decays isotropically. In contrast the $Z$ pairs from the continuum
are forward directed. Thus any uncertainty in the continuum (due to QCD
corrections and any other
parton processes contributing to the final result) will have less of an
effect on a transverse mass plot than on an invariant mass plot.

Unfortunately, in the case of
$pp\rightarrow ZZ\rightarrow\ell^+\ell^-\nu\bar{\nu}$ there are QCD background
processes\cite{Cahnetal}
which mimic the final state. At the partonic level they come from
$q\bar{q}\rightarrow Zg$ and $qg\rightarrow qZ$ where some final state
hadrons disappear down the beam pipe.
In order to remove this background it is
necessary to impose some stringent kinematic cuts. This reduces the
total number of events to around 50, and we are again facing the
problem of statistics. It is for this reason that people have been seeking
methods of enhancing the Higgs signal by using as much of the information from
the events as possible. One such method currently under investigation
is that of tagging the jets\cite{jettag}
arising from the scattered hard partons.
With appropriate cuts this effectively reduces the background to
$pp\rightarrow H\rightarrow ZZ$.

In this report we would like to suggest another method which can either
complement or supplement other methods of enhancement. It relies on the
fact that in  Higgs decay, as seen in its rest frame, the $Z$s are
predominantly longitudinally polarized whereas all other processes yield
transversely polarized $Z$s. By extracting as much information about the
production polarization of a $Z$ from the angular distribution of its decay to
the final state charged leptons it should be possible to enhance
the Higgs signal. As we shall see there is a kinematic variable
available to us in the LAB frame which allows us to do just that.

In the $Z$ pair center of mass frame we can define the fraction
of $Z$s that are longitudinally polarized,
\begin{equation}
f_L= {d\sigma_{LL}+\half d\sigma_{LT}\over d\sigma_{\rm tot}}
\end{equation}
where the subscripts refer to the polarizations of the $Z$ pair. In Fig. 2
we plot this as a function of $m_{ZZ}$ and $m_T$ for two different Higgs
masses. As we can see there is a larger area under the curve when plotted
against the transverse mass for the same reasons as discussed above.

It is important to stress that this variable can only be extracted from the
data
if we can reconstruct the $ZZ$ center of mass frame. This is because
back-to-back $Z$s have a common axis for the measurement of their spin.
Consider one of the $Z$s in this frame
and boost back along its direction of motion to its
rest frame. The longitudinal polarization vector in this `decay' frame is
$s^\mu_L=(0,{\bf n})$ where ${\bf n}$ is a unit vector
pointing along the original direction of motion.
In the $ZZ$ frame its components would be
$s^{\prime\mu}_L$ given by a simple boost from the decay to $ZZ$ frame.
In the decay frame the leptons will be back-to-back. The angle between
a lepton momentum and the unit vector is given by,
\begin{equation}
z=\cos\theta=-{2\over m_Z}(p_\ell\cdot s_L)
\end{equation}
as depicted in Fig. 3. The distribution of this angle depends
on the polarization of the decaying $Z$,
\begin{eqnarray}
\displaystyle\phi_L(z)&=&{3\over 4}\left( 1-z^2\right)\nonumber\\
\displaystyle\phi_T(z)&=&{3\over 8}\left(1+z^2\right)
\end{eqnarray}
If we can determine this angle then its distribution will yield an estimate
for $f_L$. This is an approach which has been adopted before for
$WW$ final states\cite{DKR}, and for the gold-plated mode\cite{Duncan}.

However, as we are considering one of the $Z$s decaying invisibly to neutrinos
we cannot reconstruct the $ZZ$ center of mass frame. Thus we cannot boost
all our four-momenta from the LAB to the $ZZ$ frame and thence to the decay
frame to define $s_L^\mu$. What one means by polarization is frame dependent
and so $f_L$ and $z$ cannot be extracted from the data.
Nevertheless, we can construct the longitudinal polarization vector in the LAB
frame given the reconstructed four-momentum of the decaying $Z$,
\begin{equation}
\varepsilon_L^\mu={1\over m_Z} \left( |{\bf p}_Z|,{E_Z\over |{\bf p}_Z|} {\bf
p}_Z
\right)
\end{equation}
Given this, we can define a new kinematic variable in the LAB frame,
\begin{equation}
z^\ast={2\over m_Z}\left| p_\ell\cdot \varepsilon_L\right|
\end{equation}
This will only equal $z$ when the $ZZ$ and LAB frames coincide\footnote{
This can easily be seen by considering the $Z$s being produced at
$90^0$ in the $ZZ$ frame. ${\bf s}_L$ will be perpendicular
to the beam axis as will
${\bf s}_L^\prime$. This will remain perpendicular under boosts along
the beam axis.
However, such boosts will result in the $Z$ having an angle to
the beam axis other than $90^0$, and hence $\mbox{\boldmath$\varepsilon$}_L$
will
not be perpendicular. If we were then to boost back along the $Z$ direction
$\mbox{\boldmath$\varepsilon$}_L$ would retain its direction.
Thus one cannot go directly from the LAB to the decay frame.
This is a consequence of the fact that boosts and rotations do not commute.}
{\em i.e.} when ${\bf p}_T=-\not\!{\bf p}_T$ and rapidity $y=0$. However,
since a heavy Higgs is expected to be produced almost at rest then $z^\ast$
ought to be close to $z$ when we are sitting on the resonance.

In Fig. 4 we plot some typical $z^\ast$ distributions close to and far from
a resonance. We see that in the former the distribution seems to have a lot
of $\phi_L$ in it whilst the latter has more $\phi_T$. Nevertheless, these
are not sufficiently noise-free to extract a value for $f_L^\ast$ with
small enough errors by fitting to $\phi(z^\ast)=f^\ast_L\phi_L(z^\ast)
+(1-f^\ast_L)\phi_T(z^\ast)$.
It is much more propitious to take such data and estimate the average
$\langle z^\ast\rangle$, as this should smooth out the noise a little.
This should also be of benefit if there is an error in $z^\ast$
arising from detector resolution. Theoretically we know that
$\langle z^\ast\rangle$ falls in the range,
\begin{equation}
0.375 {\rm (L)} < \langle z^\ast\rangle < 0.5625 {\rm (T)}
\end{equation}
the lower bound being for purely longitudinal $Z$ and the upper
for purely transverse $Z$. In Fig. 5 we plot $\langle z^\ast\rangle$ for
different values of the Higgs mass.
We now turn to the question of error bars.

The results presented here are preliminary. So far we have only concentrated
on the $q\bar{q}\rightarrow ZZ$ continuum for the background and the
$gg\rightarrow H\rightarrow ZZ$ for the signal. We have yet to incorporate
the hard QCD background and vector boson scattering for the signal. The
processes we have considered so far balance transverse momentum. The
vector boson scattering $qq\rightarrow qqVV\rightarrow qqZZ$ will not and
will have some effect on the distribution of $z^\ast$. This is currently
under investigation and shall report on the results soon. To get a sense
of whether this enhancement method will ultimately be of use we have plotted
$\langle z^\ast\rangle$ for two different Higgs masses in Fig. 6.
The horizontal error bars are the $m_T$ binning whilst the
vertical are estimated by varying the continuum background by $\pm 30\%$.
We see that the continuum uncertainty does not much effect
$\langle z^\ast\rangle$. This gives us hope.

The central idea of this report is that the polarization information from the
decay of the $Z$ to charged leptons can be used to help enhance the signal
over the background. We may use this information to allow us to relax
the cuts on other kinematic variables such as rapidity, transverse momenta,
{\em etc.} Doing so should leave us with a larger sample of events and thus
better statistics, particularly at higher masses. Thus by extracting
as much from the data as possible we can overcome the limitations
of a paucity of events.

\newpage
\typeout{*** Figures ***}
\begin{center}
{\bf\large Figure Captions}
\end{center}
\begin{trivlist}
\item[Fig. 1] Differential cross section for  $pp\rightarrow ZZ$ as a function
of the invariant mass of the $Z$ pair $m_{ZZ}$ (left), and as a function of the
transverse mass $m_T$ (right). For simplicity we have only considered two
contributions.
The dotted line is the $q\bar{q}\rightarrow ZZ$
continuum, the dashed line is the gluon fusion $gg\rightarrow H\rightarrow ZZ$
Higgs production. The solid line is the sum of both.
They are calculated for $m_H=400$ GeV and $m_t=150$ GeV.
\item[Fig. 2] Longitudinal fraction (as defined in the $ZZ$ center of mass
frame)
plotted as a function of the invariant mass $m_{ZZ}$ (left) and the transverse
mass $m_T$ (right) for Higgs masses 400 GeV (upper) and 800 GeV (lower).
We have included only the
$q\bar{q}\rightarrow ZZ$ continuum and the $gg\rightarrow H\rightarrow ZZ$
parton subprocesses for demonstration.
\item[Fig. 3] Relationship between the `decay', $ZZ$, and LAB frames. The decay
frame
can only be reached from the LAB via the $ZZ$ frame. This is because the
polarization vector is only covariant under boosts along the direction of
motion of the $Z$.
\item[Fig. 4] The frequency of $z^\ast$ in each of 20 bins. The left histogram
is for all events in a 50 GeV $m_T$ bin close to a 400 GeV resonance.
The right histogram is for a 50 GeV bin far from a 800 GeV resonance.
\item[Fig. 5] The average value of $z^\ast$ for 50 GeV $m_T$ bins. The top
curve is where the Higgs contribution is not included. The next curve is
for a Higgs of 1200 GeV. The other three curves are for Higgses of
400, 600 and 800 GeV respectively.
\item[Fig 6] The average $\langle z^\ast\rangle$ in a particular $m_T$ bin
for a 400 GeV Higgs (left) and a 800 GeV Higgs (right). The triangles
are for pure continuum and we see that the results sit around the
upper bound of Eq. (7). The error bars on the signal+background runs
are obtained by varying the continuum by $\pm 30\%$. We have not included
the hard QCD background, imposed cuts, nor estimated the statistical errors.
\end{trivlist}

\begin{thebibliography}{10}
\typeout{*** Citations ***}
\bibitem{Cahn}R.N.Cahn and M.S.Chanowitz, \prl {\bf 56} (1986) 1327.
\bibitem{Cahnetal}R.N.Cahn, M.Chanowitz, M.Gilchriese, M.Golden, J.Gunion,
M.Herrero, I.Hinchliffe, F.Paige and E.Wang,
in {\em Experiments, Detectors, and Experimental Areas for the SSC},
proceedings of the Workshop, Berkeley, CA, 1987, {\em eds.} R.Donaldson
and M.Gilchriese (World Scientific, Singapore, 1988).
\bibitem{jettag}V.Barger, T.Han and R.J.N.Phillips, \pr {\bf D37}
(1988) 2005;\\
V.Barger, K.Cheung, T.Han, J.Ohnemus and D.Zeppenfeld, \pr {\bf D44}
(1991) 1426;\\
V.Barger, K.Cheung, T.Han and D.Zeppenfeld, \pr {\bf D44} (1991) 2701;\\
V.Barger, K.Cheung, T.Han and D.Zeppenfeld, preprint MAD/PH/757
(Madison, WI, May 1993);\\
J.Bagger, V.Barger, K.Cheung, J.Gunion, T.Han, G.A.Ladinsky, R.Rosenfeld
and C.P.Yuan, preprint Fermilab-Pub-93-040-T
(Batavia, IL, June 1993).
\bibitem{DKR}M.J.Duncan, G.L.Kane and W.W.Repko, \np {\bf B272}
(1986) 517;\\
G.L.Kane and C.P.Yuan, \pr {\bf D40} (1989) 2231.
\bibitem{Duncan}M.J.Duncan, \plt {\bf B179} (1986) 393;\\
T.Matsuura and J.J.Van der Bij, {\it Z. Phys. } {\bf C51} (1991) 259.
\end{thebibliography}
\end{document}